\documentclass[onecolumn,prd,nofootinbib,aps,floats,floatfix,amsmath,amssymb,secnumarabic,preprintnumbers,showpacs,notitlepage]{revtex4-1} %
\usepackage[final]{graphicx}
\usepackage{ulem}
\usepackage{amsmath}

\def\beq{\begin{equation}}
\def\eeq{\end{equation}}
\def\bea{\begin{eqnarray}}
\def\eea{\end{eqnarray}}
\def\nn{\nonumber}

\def\roughly#1{\mathrel{\raise.3ex\hbox
{$#1$\kern-.75em\lower1ex\hbox{$\sim$}}}}
\def\lsim{\roughly<}

\def\sla#1{\raise.15ex\hbox{$/$}\kern-.57em #1}% Feynman slash

\allowdisplaybreaks

\begin{document}
\rightline{UdeM-GPP-TH-20-276} 
\title{\boldmath CP Violation in Same-sign Dilepton Production at the LHC}

\author{Fatemeh Najafi}\email{fatemeh.najafi@umontreal.ca}
\affiliation{Physique des Particules, Universit\'e de Montr\'eal,
C.P. 6128, succ. centre-ville, Montr\'eal, QC H3C 3J7, Canada}
\author{Jacky Kumar}\email{jacky.kumar@umontreal.ca}
\affiliation{Physique des Particules, Universit\'e de Montr\'eal,
C.P. 6128, succ. centre-ville, Montr\'eal, QC H3C 3J7, Canada}
\author{David London}\email{london@lps.umontreal.ca}
\affiliation{Physique des Particules, Universit\'e de Montr\'eal,
C.P. 6128, succ. centre-ville, Montr\'eal, QC H3C 3J7, Canada}
\date{\today}
\author{Richard MacKenzie}\email{richard.mackenzie@umontreal.ca}
\affiliation{Physique des Particules, Universit\'e de Montr\'eal,
C.P. 6128, succ. centre-ville, Montr\'eal, QC H3C 3J7, Canada}

\begin{abstract}
If the neutrino is a Majorana particle, low-energy
lepton-number-violating (LNV) processes, such as neutrinoless
double-beta ($0\nu\beta\beta$) decay, are possible. It may also be
possible to observe high-energy $0\nu\beta\beta$-like LNV processes at
the LHC. These are distinguished by the presence of same-sign
dileptons in the final state (e.g., ${\bar u} d \to t {\bar b} \, e^-
\mu^-$). In this paper, we show that CP-violating triple products
(TPs) may be present in the process, and may be measurable at the
LHC. If a nonzero TP were observed, it would give us much information
about the underlying new physics (NP). We would know that there are
(at least) two interfering NP amplitudes, with different weak phases
and different Lorentz structures. And if we had some knowledge of the
NP, e.g., by direct production of NP particles, we could get
information about the magnitudes and relative phases of its couplings.
\end{abstract}

%\pacs{11.30.Er,13.25.Hw, 12.60.-i}
\maketitle

\section{Introduction}

One of the outstanding questions in particle physics is the nature of
the neutrino. In particular, is it a Majorana particle? If it is, then
lepton-number-violating processes, such as neutrinoless double-beta
($0\nu\beta\beta$) decay, are possible. A great deal of time and
effort has been spent looking for $0\nu\beta\beta$ decay, but to date,
no signal has been seen (for a review, see
Ref.~\cite{Dolinski:2019nrj}).

The $0\nu\beta\beta$ process is $n n \to p p \, e^- e^-$, which at the
quark level is $d d \to u u \, e^- e^-$. While $0\nu\beta\beta$ decay
is a low-energy process, $d d \to u u \, e^- e^-$ could, in principle,
also be observable at the LHC, given that $p p$ collisions are
involved. Furthermore, as this would now be a high-energy process, one
or both of the final-state $e^-$'s could be a $\mu^-$ or a
$\tau^-$. So not only is the process lepton-number-violating, it could
also be lepton-flavor-violating: $d d \to u u \, \ell^- \ell^{\prime
  -}$. In addition, $p p$ collisions will also generate the related
processes $d {\bar u} \to u {\bar d} \, \ell^- \ell^{\prime -}$ and
${\bar u} {\bar u} \to {\bar d} {\bar d} \, \ell^- \ell^{\prime -}$,
as well as their CP conjugates.  Finally, the $d$ and $u$ quarks can
be down-type and up-type quarks of any family. Thus, what is studied
at the LHC is really many processes: $d_i d_j \to u_k u_l \, \ell^-
\ell^{\prime -}$, $d_i {\bar u}_j \to u_k {\bar d}_l \, \ell^-
\ell^{\prime -}$, ${\bar u}_i {\bar u}_j \to {\bar d}_k {\bar d}_l \,
\ell^- \ell^{\prime -}$. We refer to all of these as
$0\nu\beta\beta$-like processes, which are identified by the presence
of same-sign dileptons in the final state.

On the other hand, there is also a huge disadvantage at the LHC.
Assuming that the neutrino masses are generated via the seesaw
mechanism, the three ultra-light neutrinos are Majorana (leading to
lepton number violation) and mix among themselves (leading to lepton
flavor violation).  There are also three heavy neutrinos, which have
little effect at low energy. The key point is that, in the standard
model (SM) with Majorana neutrinos, the diagram for the process $d d
\to u u \, e^- e^-$ involves a neutrino propagator, with the result
that the amplitude is proportional to $m_\nu$, which is tiny,
$O(10^{-2})$ eV. Low-energy experiments are approaching the
sensitivity to observe $0\nu\beta\beta$ decay, even with such a
suppression factor. However, if the amplitude is, in fact,
proportional to $m_\nu$, the process would be completely unobservable
at the LHC.

Even so, there are many new-physics (NP) models in which
$0\nu\beta\beta$-like processes can be generated {\it without} the
amplitude being suppressed by a light neutrino mass. (For a review of
NP models that can contribute to $0\nu\beta\beta$ decay, see
Ref.~\cite{Deppisch:2012nb}.) If a $0\nu\beta\beta$-like process were
observed at the LHC, it would point to the presence of one of these
models of NP. The question is: which one?

Ultimately, this question can only be answered by the direct
production of the NP particles themselves.  In this paper, we show
that there are indirect ways of learning about the NP: (i)
$0\nu\beta\beta$-like processes potentially include CP-violating
observables (triple products), and (ii) the measurement of such CP
violation would give us important information about the underlying NP
that cannot be easily obtained otherwise.

We begin in Sec.~2 with an effective-field-theory (EFT) analysis of
$0\nu\beta\beta$-like processes. We find that the interference of
certain operators can give rise to CP-violating triple producs. We
illustrate this explicitly in Sec.~3 by constructing a toy model that
produces such CP-violating effects. We show how the measurement of
non-zero triple producs can give us information about this model. In
Sec.~4, we explore the experimental prospects for making such
measurements at the LHC. We conclude in Sec.~5.

\section{EFT Analysis}

CP violation arises due to the interference of (at least) two
amplitudes with a relative weak (CP-odd) phase.  The first step is
therefore to identify all possible operators that can contribute to
this process.

To this end, following the notation of Ref.~\cite{Bonnet:2012kh}, we
list all the dimension-9 operators that contribute to $d_i d_j \to u_k
u_l \, \ell^- \ell^{\prime -}$ and the related $0\nu\beta\beta$-like
processes. These operators take the form ${\bar u} \Gamma_1 d \, {\bar
  u} \Gamma_2 d \, {\bar \ell}' \Gamma_3 \ell^C$, where the $\Gamma_i$
include all possible Lorentz structures (detailed below). Here we have
suppressed the flavor indices, so that $u$, $d$ and $\ell, \ell'$
represent any of $\{u, c, t\}$, $\{d, s, b\}$ and $\{e, \mu, \tau\}$,
respectively. All operators involve two hadronic currents $J$ and one
leptonic current $j$. Each of these has three types of Lorentz
structure:
\begin{align}
%\label{hello}
J_{L,R} &\equiv \bar u P_{L,R} \, d ~,  & j_{L,R} &\equiv {\bar \ell}' P_{L,R} \, \ell^C ~, \nonumber \\
J^\mu_{L,R} &\equiv \bar u \, \gamma^\mu P_{L,R} \, d ~, & j^\mu_{L,R} &\equiv {\bar \ell}' \gamma^\mu P_{L,R} \, \ell^C ~, \\
J^{\mu\nu}_{L,R} &\equiv \bar u \, \sigma^{\mu\nu} P_{L,R} \, d ~ ~, & j^{\mu\nu}_{L,R}
&\equiv {\bar \ell}' \sigma^{\mu\nu} P_{L,R} \, \ell^C ~, \nonumber
\end{align}
%\bea
%& J_{L,R} \equiv {\bar u} P_{L,R}d ~~,~~~~ J_{L,R}^{\mu} \equiv {\bar u}\gamma^\mu P_{L,R}d ~, & \nn\\
%& J_{L,R}^{\mu\nu} \equiv {\bar u} \sigma^{\mu \nu} P_{L,R}d ~, & \nn\\ 
%& j_{L,R} \equiv {\bar \ell}' P_{L,R}\ell^C ~~,~~~~  j_{L,R}^{\mu} \equiv {\bar \ell}'\gamma^\mu P_{L,R} \ell^C ~, & \nn\\
%& j_{L,R}^{\mu\nu} \equiv {\bar \ell}' \sigma^{\mu \nu} P_{L,R} \ell^C ~, &
%\eea
%
where the antisymmetric tensor is defined as $\sigma^{\mu \nu} =
\frac{\emph{i}}{2} [\gamma^\mu, \gamma^\nu]$. 

Note that, if $\ell = \ell'$, the leptonic current must be
antisymmetric under the exchange of the two identical leptons. This
implies that
\beq 
\label{eq:ee}
{\bar \ell} \gamma^\mu \ell^C =0 ~,~~ {\bar \ell} \sigma^{\mu \nu} \ell^C =0 ~.
\eeq

For simplicity, we consider only dimension-9 operators involving
hadronic currents that are colour singlets. The most general effective
Lagrangian containing such dimension-9 operators is then given by
\begin{equation}
\mathcal{L}_{\text eff} = \frac{1}{M^5}\sum_{X,Y,Z=L,R} \,\ \sum_{i=1}^8 C_i^{(XY)Z} \, (O_i)_{(XY)Z} \,\ + h.c.,
\end{equation}
where $M$ is the scale of NP, and
\bea
{\rm SSS:}~~~~(O_1)_{(XY)Z} &=& J_X J_Y j_Z ~, \nn\\
{\rm TTS:}~~~~(O_2)_{(XY)Z}  &=& (J_X)_{\mu \nu} (J_Y)^{\mu \nu} j_Z ~,   \nn\\ 
{\rm VVS:}~~~~(O_3)_{(XY)Z} &=& (J_X)_\mu (J_Y)^\mu j_Z ~, \nn\\
{\rm TVV:}~~~~(O_4)_{(XY)Z}  &=& i (J_X)_{\mu\nu} (J_Y)^\mu (j_Z)^\nu ~,  \nn\\ 
{\rm SVV:}~~~~(O_5)_{(XY)Z} &=& J_X (J_Y)_\mu (j_Z)^\mu ~, \\ 
{\rm VVT:}~~~~(O_6)_{(XY)Z} &=&  i(J_X)_\mu (J_Y)_\nu (j_Z)^{\mu\nu} ~, \nn\\ 
{\rm STT:}~~~~(O_7)_{(XY)Z} &=& (J_X)_{\mu \nu} J_Y (j_Z)^{\mu \nu} ~, \nn\\
{\rm TTT:}~~~~(O_8)_{(XY)Z} &=& i(J_X)_{\mu \mu^\prime} (J_Y)^{\mu \nu^\prime} (j_Z)^{\mu^\prime}_{\nu^\prime} ~. \nn
\label{dim9ops}
\eea
Here we denote the scalar, vector and tensor currents as S, V and T,
respectively. With this shorthand, we describe each operator as a
product of these different Lorentz structures. For example, in the
third entry, $O_3$ is VVS, where the first two labels (V) denote the
hadronic currents, and the third (S) is the leptonic current.
Furthermore, since the first two currents are both hadronic, the
labels should be understood as being symmetric in these currents. That
is, $O_4$ is both TVV and VTV, and similarly for operators $O_5$ and
$O_7$. The operators $O_{4,6,8}$, which involve an odd number of
tensor currents, include a prefactor of $i$ to compensate the factor
of $i$ in $\sigma_{\mu \nu}$.  The operators $O_6$-$O_8$ are nonzero
only if $\ell \ne \ell'$ [see Eq.~(\ref{eq:ee})].

Note that the above operators are written at the level of weak
effective theory (WET), i.e., after electroweak symmetry breaking.
That is, they are invariant only under $SU(3)_C \times U(1)_{\rm em}$.
When considering a particular operator, one must ensure that it is
compatible with SMEFT (the SM effective field theory), i.e., it
respects the full $SU(3)_C \times SU(2)_L \times U(1)_Y$ gauge
symmetry. Only a subset of the above dimension-9 operators are already
part of the SMEFT. The others must be generated in other ways, e.g.,
by dimension-11 operators containing the Higgs. The dimension-9
operator is obtained when the Higgs gets a vev
\cite{Jordy,Babu:2001ex}.

As noted above, we propose to obtain information about the underlying
NP through the measurement of CP-violating observables in this decay.
These observables arise due to the interference of two of the above
operators. In our analysis, we neglect the masses of all fermions,
except for that of the top quark.  Now, the interference of the
left-handed and right-handed fermion fields $f_L$ and $f_R$ is
proportional to $m_f$, so that it vanishes in the limit $m_f \to
0$. This implies that, in the two interfering amplitudes, each fermion
field in one amplitude must have the same chirality as the
corresponding fermion field in the other amplitude. (The only
exception is if the final state includes two top quarks.) Clearly each
current can interfere with another current of the same Lorentz
structure. However, if we consider two different types, only S-T
interference is allowed. The key point here is that, since only S-S,
V-V, T-T and S-T interferences are allowed, we can immediately see
which operators interfere and which do not. For example, $O_1$ and
$O_2$ interfere, but $O_1$ and $O_3$ do not.

\section{Toy Model}

Now, in Eq.~(\ref{dim9ops}) there are several pairs of operators that
can interfere: SSS-TTS, VVS-VVT, etc., and each pair has its own set
of CP-violating effects. Furthermore, these effects depend on which
$0\nu\beta\beta$-like process is used. In this paper, in order to
clearly illustrate the various features of our method, we focus on a
single pair of operators -- SSS and STT -- and examine the
$0\nu\beta\beta$-like process ${\bar u} d \to t {\bar b} \, e^-
\mu^-$, in which there are no identical particles.  In this section,
we construct a toy model to generate the SSS and STT operators. Note
that we are not advocating this model; it is chosen only for
illustrative purposes.

\begin{figure}[!htbp]
\begin{center}
\includegraphics[width=0.3\textwidth]{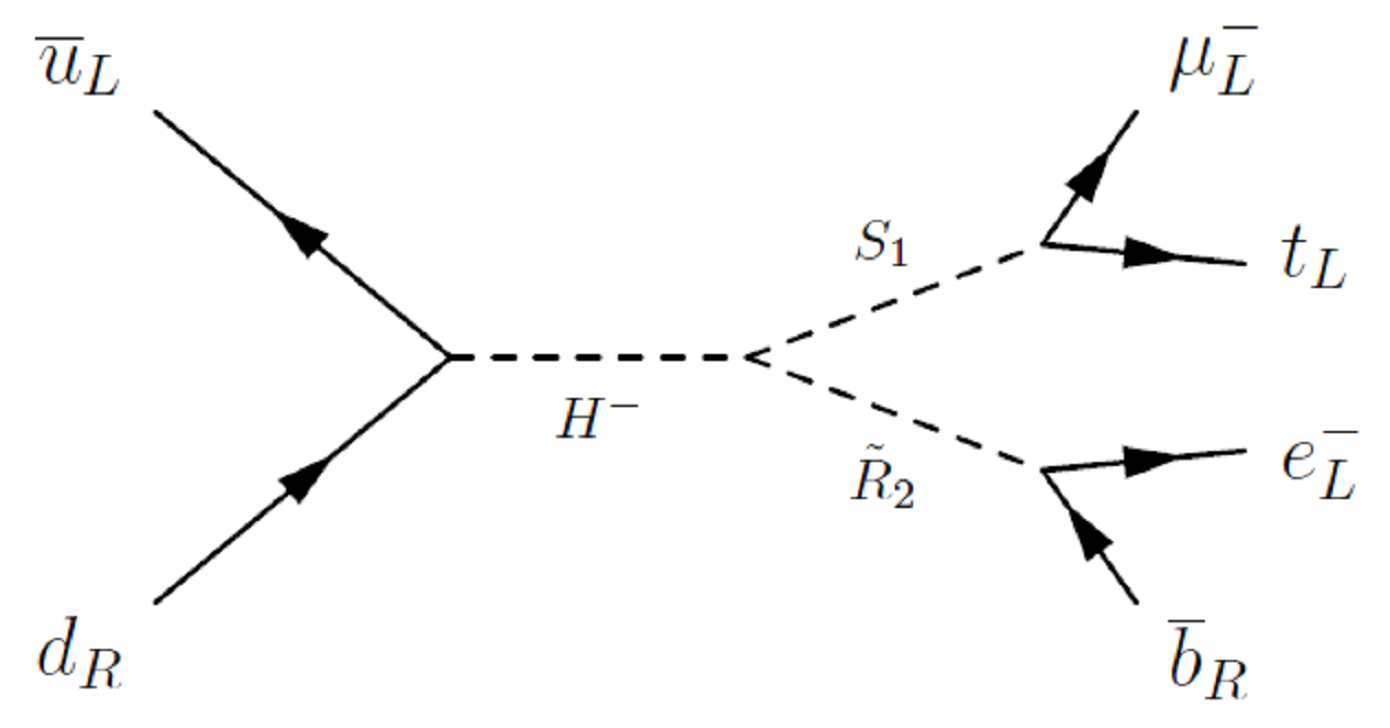}
\end{center}
\caption{\small Contribution to ${\bar u} d \to t {\bar b} \, e^-
  \mu^-$ involving no virtual neutrinos.}
\label{HLQLQfig}
\end{figure}

One question that may arise at this stage is: assuming that the NP
particles are scalars, fermions or vectors, how can there be tensor
operators? The answer is that these can be generated via Fierz
transformations. As an example of how this can come about, suppose
that ${\bar u} d \to t {\bar b} \, e^- \mu^-$ is produced as
follows. We have ${\bar u} d \to H^-$, where $H^-$ is a charged
(scalar) Higgs boson, part of an $SU(2)_L$ doublet with $Y = 1/2$.  We
also have two scalar leptoquarks (LQs), ${\tilde R}_2$ and $S_1$
\cite{LQs}, that decay as follows: ${\tilde R}_2 \to {\bar b}_R e_L^-$
(fermion-number conserving) and $S_1 \to t_L \mu_L^-$ (fermion-number
violating).  Finally, we allow $H^- \to {\tilde R}_2 S_1$. This
coupling conserves all SM quantum numbers, but it violates lepton
number by 2 units. Thus, all couplings respect the full $SU(3)_C
\times SU(2)_L \times U(1)_Y$ symmetry, so that this model is
compatible with the SMEFT. The diagram of this process is shown in
Fig.~\ref{HLQLQfig}.

Note that, since the model generates $\Delta L=2$
$0\nu\beta\beta$-like processes, it may also contribute to Majorana
neutrino-mass terms at higher order (after electroweak symmetry
breaking). In the present case, if the $H^-$ also couples to ${\bar
  t}b$, it will produce a neutrino mass term at two loops. Since the
present limits on neutrino masses are at the scale of $O({\rm eV})$,
this could be problematic: given that the scale of NP is $O({\rm
  TeV})$, a two-loop suppression could still lead to a value of
$m_\nu$ that is many orders of magnitude too large. Fortunately, here
the problem can be evaded simply by taking the $H^- {\bar t}b$
coupling $\simeq 0$, but these types of potential problems should be
checked, even for a toy model.

Referring to Fig.~\ref{HLQLQfig}, when the heavy NP particles are
integrated out, one obtains the dimension-9 operator
\beq
\frac{M'}{M^6} \, {\bar u} P_R d \, {\bar e} P_R b \, {\bar t} P_R \mu^C ~.
\label{model}
\eeq
The prefactor $M'/M^6$ arises from two sources. First, the
$H^-$-${\tilde R}_2$-$S_1$ coupling is proportional to a mass,
$M'$. Second, the propagator of each of $H^-$, ${\tilde R}_2$ and
$S_1$ provides a factor $1/M^2_{\text part}$. Taking the masses of all
virtual particles to be the same size, one arrives at a prefactor
$M'/M^6$. In order to maximize the effect of this contribution, we
take $M$ to be as small as possible, given the present experimental
limits from direct searches. This means that $M=O({\rm TeV})$.

Performing a Fierz transformation of Eq.~(\ref{model}), one obtains
\beq
\frac{M'}{M^6} \left[ {1\over 2} {\bar u} P_R d \, {\bar t} P_R b \, {\bar e} P_R \mu^C 
+ {1\over 8} {\bar u} P_R d \, {\bar t} \sigma_{\mu\nu} P_R b \, {\bar e} \sigma^{\mu\nu} P_R \mu^C \right] ~.
\label{modelFierz}
\eeq
Thus, with only scalar NP particles, this model produces both SSS
($O_1$) and STT ($O_7$) operators. 

Note that, since these two operators have the same weak phase, their
interference does not generate CP violation. (This is obvious, since
there is basically only a single operator, Eq.~(\ref{model}).) In
order to produce a CP-violating effect, we must interfere two
operators with different weak phases. This is done below.

Turning to CP violation, the most common CP-violating observable is
the direct CP asymmetry, which is the difference in the rates of the
process and the CP-conjugate process. A nonzero direct CP asymmetry
requires not only a weak-phase difference between the two interfering
amplitudes, but also a strong-phase difference. In the present case,
if the two interfering amplitudes were, for example, VVS and VVT, the
two amplitudes would have the same hadronic structures. We would
therefore expect the strong phases to also be the same, resulting in a
vanishing direct CP asymmetry. And even with SSS-STT interference,
although the Lorentz structures are different, the QCD structure
(i.e., the placement of the quark fields) is the same in the two
amplitudes, so that the strong phases should be similar.  The upshot
is that we do not expect a sizeable direct CP asymmetry in $d_i d_j
\to u_k u_l \, \ell^- \ell^{\prime -}$.\footnote{A possible loophole
  is if one of the interfering amplitudes involves a resonant
  decay. In this case, the strong phase can be generated by the
  absorptive part of the amplitude, see Ref.~\cite{Bray:2007ru}.}

Another type of CP-violating observable involves triple product (TP)
correlations \cite{TPs1, TPs2}. These take the form ${\vec v}_1 \cdot
({\vec v}_2 \times {\vec v}_3)$, where the ${\vec v}_i$ are momenta or
polarizations. Technically, while the TP is T-odd, it is not
CP-violating, as it can be generated by strong phases. A true
CP-violating observable can be obtained by comparing the TPs in a
process and its CP-conjugate process.  However, if the strong phases
are negligible, as is expected here, then a nonzero TP in a single
process is an indication of CP violation.

To illustrate how TPs can arise, we return to the model above
[Eqs.~(\ref{model}) and (\ref{modelFierz})]. Suppose that ${\tilde
  R}_2$ has two decay modes: ${\tilde R}_2 \to {\bar b}_R e_L^-$ and
${\tilde R}_2 \to {\bar b}_R \mu_L^-$, with different (complex)
couplings. Similarly, $S_1 \to t_L \mu_L^-$ and $S_1 \to t_L e_L^-$,
also with different couplings. There are now two amplitudes
contributing to ${\bar u} d \to t {\bar b} \, e^- \mu^-$:
\bea
\label{op1}
(i) &:& A_1 = c_1 \, \frac{M'}{M^6} \, {\bar u} P_R d \, {\bar e} P_R b \, {\bar t} P_R \mu^C ~, \\
\label{op2}
(ii) &:& A_2 = c_2 \, \frac{M'}{M^6} \, {\bar u} P_R d \, {\bar \mu} P_R b \, {\bar t} P_R e^C ~.
\eea
The coefficients $c_1$ and $c_2$ are each products of four couplings:
\beq
c_1 = c_H^{{\bar u}d} \, c_H^{{\tilde R}_2 S_1} \, c_{{\tilde R}_2}^{{\bar b}e} \, c_{S_1}^{t \mu} ~~,~~~~
c_2 = c_H^{{\bar u}d} \, c_H^{{\tilde R}_2 S_1} \, c_{{\tilde R}_2}^{{\bar b}\mu} \, c_{S_1}^{t e} ~,
\label{c1c2}
\eeq
where $c_P^{ij}$ is the coupling of the scalar $P$ ($H^-$, ${\tilde
  R}_2$ or $S_1$) to particles $i$ and $j$.

The total amplitude is the sum of these two amplitudes: $A_{\text tot} = A_1
+ A_2$. When we compute $|A_{\text tot}|^2$, these two interfere. In the
interference of Eqs.~(\ref{op1}) and (\ref{op2}), one finds a term of
the form
\bea
&& {\rm Re} \left[ (c_1 c_2^*) \, {\rm Tr}[\sla{p}_{{\bar u}} \sla{p}_d] \, 
{\rm Tr} [ \sla{p}_e \sla{p}_{{\bar b}} \sla{p}_\mu \sla{p}_t \gamma_5 ] \right] \nn\\
&& \hskip10truemm \propto ~
{\rm Im}(c_1 c_2^*) \, p_{{\bar u}} \cdot p_d \, \epsilon_{\mu \nu \rho \sigma} \, p_e^\mu p_{{\bar b}}^\nu p_\mu^\rho p_t^\sigma ~.
\label{TPterm}
\eea
This is a TP term. The 4-momenta of each of the final-state
particles can be measured, so that $\epsilon_{\mu \nu \rho \sigma} \,
p_e^\mu p_{{\bar b}}^\nu p_\mu^\rho p_t^\sigma$ includes four
different TPs: $E_t \, {\vec p}_{{\bar b}} \cdot ({\vec p}_e \times
{\vec p}_\mu)$, $E_{\bar b} \, {\vec p}_t \cdot ({\vec p}_e \times
{\vec p}_\mu)$, $E_e \, {\vec p}_{{\bar b}} \cdot ({\vec p}_t \times
{\vec p}_\mu)$, $E_\mu \, {\vec p}_{{\bar b}} \cdot ({\vec p}_e \times
{\vec p}_t)$. However, individually these terms are not
Lorentz-invariant. It is only the original term, $p_{{\bar u}} \cdot
p_d \epsilon_{\mu \nu \rho \sigma} \, p_e^\mu p_{{\bar b}}^\nu
p_\mu^\rho p_t^\sigma$, that is Lorentz-invariant. From here on, we
refer to this as the Lorentz-invariant triple product (LITP).

Using this, one can now construct the TP asymmetry. For each event,
the LITP is computed. This information can then be used to obtain
\beq
A_{TP} = \frac{ {\hbox{\# events (LITP $>0$)}} - {\hbox{\# events (LITP $<0$)}}}{{\hbox{total \# events}}} ~.
\label{ATP}
\eeq
If $A_{TP} \ne 0$, this indicates a nonzero LITP, which is a signal of
CP violation.

\section{Experimental Prospects}

At this stage, the question is: could this be measured at the LHC? To
explore this, we implemented the model in {\tt FeynRules}
\cite{Alloul:2013bka} and used {\tt MadGraph} \cite{Alwall:2014hca} to
generate events. We considered three versions of the LHC: (i) the
high-luminosity LHC (HL-LHC, $\sqrt{s} = 14$ TeV, peak ${\cal L} =
3~{\rm ab}^{-1}$), (ii) the high-energy LHC (HE-LHC, $\sqrt{s} = 27$
TeV, peak ${\cal L} = 15~{\rm ab}^{-1}$) \cite{Zimmermann:2017bbr},
(iii) the future circular collider (FCC-hh, $\sqrt{s} = 100$ TeV, peak
${\cal L} = 30~{\rm ab}^{-1}$) \cite{Golling:2016mxw}.

We fix the parameters of the model. We take the $H{\bar u}d$ coupling
to be $|c_H^{{\bar u}d}| = 0.1$ and the LQ couplings to be
$|c_{{\tilde R}_2}^{{\bar b}e}| = |c_{S_1}^{t \mu}| = |c_{{\tilde
    R}_2}^{{\bar b}\mu}| = |c_{S_1}^{t e}| = 1$, with the relative
weak phase of $c_1$ and $c_2$ [Eq.~(\ref{c1c2})] equal to $\pi/2$. We
take the $H^-$-$S_1$-${\tilde R}_2$ coupling to be $M' = 1$ TeV.

Turning to the masses of the NP particles, we must check that these
parameters respect the experimental bounds:
\begin{enumerate}

\item The charged Higgs mass is constrained by the search for dijet
  resonances. In Ref.~\cite{Sirunyan:2018xlo}, it is found that $M_H =
  1$ TeV is allowed, as long as $\sigma {\cal B} \lsim 1$ pb, where
  $\sigma$ is the $H^\pm$ production cross section, and ${\cal B}$ is
  the branching ratio of the $H^\pm$ to two jets. In our case, ${\cal
    B} = 1$, and $\sigma = 0.6$ pb for $|c_H^{{\bar u}d}| = 0.1$. So
  our value of $M_H = 1$ TeV is allowed. (Note that the dijet
  constraint is actually stronger for heavier resonances, so the data
  easily allow a lighter charged Higgs.)

\item Bounds on LQs are given in Ref.~\cite{Angelescu:2018tyl}. There
  are two sources: (i) pair production of LQs, (ii) $t$-channel
  contribution of LQs to $pp \to \ell^+ \ell^-$ ($\ell = \mu,
  \tau$). For $S_1$, it is found that its mass can be 950 GeV if its
  branching ratio to $t \mu$ is 50\% (as it is in our model).
  ${\tilde R}_2$ is not discussed in Ref.~\cite{Angelescu:2018tyl},
  but $R_2$ is. Assuming similar bounds, the mass of ${\tilde R}_2$
  can be as low as 1160 GeV if its branching ratio to ${\bar b}\mu$ is
  50\% and its coupling to ${\bar b}\mu$ is 1 (as it is in our
  model). This limit can be weakened if other decays are allowed. In
  light of all this, we take the LQ masses to be $M_{S_1} = M_{{\tilde
      R}_2} = 1$ TeV.

\end{enumerate}

In addition to the process ${\bar u} d \to t {\bar b} \, e^- \mu^-$,
there is also the CP-conjugate process, $u {\bar d} \to {\bar t} b \,
e^+ \mu^+$. The amplitude for the anti-process is obtained from that
for the process by simply changing the sign of the weak-phase
difference. Now, one can show that the TP in the CP-conjugate process
is equal to that in the process: there is a minus sign coming from the
weak phase, and another minus sign coming from the parity-odd angular
function \cite{TPs2}. So one can combine both processes in measuring
the TP asymmetry.

Using this NP model, MadGraph generates $p p \to t {\bar b} \, e^-
\mu^-$ events, giving the 4-momenta of the final-state particles
for each event. Using energy-momentum conservation, $p_{{\bar u}} +
p_d = p_e + p_{{\bar b}} + p_\mu + p_t$, so that
\beq
p_{{\bar u}} \cdot p_d = \frac12 (p_{{\bar u}} + p_d)^2 = (p_e + p_{{\bar b}} + p_\mu + p_t)^2 ~.
\eeq
With this, the LITP $p_{{\bar u}} \cdot p_d \, \epsilon_{\mu \nu \rho
  \sigma} \, p_e^\mu p_{{\bar b}}^\nu p_\mu^\rho p_t^\sigma$ is
computed. Doing this for all events, the LITP asymmetry $A_{TP}$
[Eq.~(\ref{ATP})] is calculated. This procedure is repeated for the
CP-conjugate process $p p \to {\bar t} b \, e^+ \mu^+$.

\begin{table}[h]
\begin{center}
\begin{tabular}{c|c|c|c|c }
        \hline \hline
Machine ($\sqrt{s})$    & Peak $\mathcal{L}$ & $\sigma$(fb) & Expected $\#$ events & $A_{TP}$ \\ 
\hline
HL-LHC (14 TeV)&3 $\rm ab^{-1}$  & 0.005       & 15 & $14\%$ \\
HE-LHC (27 TeV)&15 $\rm ab^{-1}$ & 0.03       & 450 & $9.2\%$  \\
FCC-hh (100 TeV)  &30 $\rm ab^{-1}$ & 0.24    & 7.2K & $5.1\%$\\
        \hline  \hline
\end{tabular}
\vskip4truemm
\begin{tabular}{c|c|c|c|c }
        \hline \hline
Machine ($\sqrt{s})$    & Peak $\mathcal{L}$ & $\sigma$(fb) & Expected $\#$ events & $A_{TP}$ \\ 
\hline
HL-LHC (14 TeV)&3 $\rm ab^{-1}$  & 0.01       & 30 & $12.3\%$ \\
HE-LHC (27 TeV)&15 $\rm ab^{-1}$ & 0.05       & 750 & $7.3\%$  \\
FCC-hh (100 TeV)  &30 $\rm ab^{-1}$ & 0.32    & 9.6K & $4.2\%$\\
        \hline  \hline
\end{tabular}
\caption{Summary for $pp\to t \bar b \, e^- \mu^-$ (top) and $pp\to
  \bar t b \, e^+ \mu^+$ (bottom).  The LITP asymmetry is calculated
  using Madgraph with a simulated sample of $10^6$ events. }
\end{center}
\label{ATPresults}
\end{table}

The results are shown in Table I. There are three patterns that should
be explained.
\begin{enumerate}

\item For each machine, the cross section for the process is smaller
  than that for the anti-process. Explanation: The process $pp\to t
  \bar b \, e^- \mu^-$ and anti-process $pp\to \bar t b \, e^+ \mu^+$
  involve ${\bar u} d$ and $u {\bar d}$ annihilation, respectively.
  But there are more $u$ quarks in a proton than $d$ quarks.

\item $A_{TP}$ decreases as the energy increases. Explanation: The TP
  is produced in SSS-STT interference, and the STT amplitude can be
  generated via a Fierz transformation of an SSS amplitude. These
  effective operators are produced by the exchange of virtual
  particles. But in some events, the final state is produced by the
  decay of (at least) one {\it on-shell}\/ LQ. In this case, there is
  no effective operator, which means no Fierz transformation, and
  hence no TP. The number of events with an on-shell LQ increases with
  increasing energy, resulting in a smaller $A_{TP}$.

\item While $A_{TP}$ is similar for process and anti-process, it is
  always a bit smaller for the anti-process. Explanation: As noted
  above, the events are generated by ${\bar u} d$ or $u {\bar d}$
  annihilation. This means that they don't all have the same
  energy. Because the proton contains two valence $u$ quarks, but only
  one valence $d$ quark, the average $u$-quark energy is a bit larger
  than the average $d$-quark energy. As a result, on average, the
  anti-process events have a larger energy, which results in more
  on-shell LQs, and hence a smaller $A_{TP}$. 

\end{enumerate}

Adding the results from both Tables, we see that the expected number
of events at the HL-LHC, HE-LHC and FCC-hh are about 45, 1200 and
17,000, respectively. Now, given an LITP asymmetry $A_{TP}$, the
number of events required to show that it is nonzero at $n\sigma$ is
\beq
N = \frac{n^2}{A_{TP}^2 \, \epsilon} ~,
\eeq
where $\epsilon$ is the experimental efficiency. Using this, we see
that (i) $A_{TP} \simeq 13\%$ is not measurable at the HL-LHC, (ii)
depending on the value of $\epsilon$, $A_{TP} \simeq 8\%$ may be
measurable at the HE-LHC at the level of $\simeq 2\sigma$, and (iii)
$A_{TP} \simeq 4.5\%$ is certainly measurable at the FCC-hh at the
$3\sigma$ level.

Of course, the above analysis is purely ``theoretical,'' i.e., it does
not take into account the issues an actual experiment will have to
deal with. A complete analysis would include a full Monte Carlo
simulation, but this is beyond the scope of this paper. Instead, we
list below several things that will have to be taken into account in a
real analysis, along with a discussion of the implications.

First, there is the question of backgrounds. Both the ATLAS and CMS
Collaborations have searched for NP at the LHC using same-sign
dilepton events, and have had to deal with backgrounds. Consulting
Ref.~\cite{Khachatryan:2016kod}, we find that the background most
relevant for us is $t {\bar t} W^-$, in which ${\bar t} \to {\bar b}
W^- (\to \ell^- {\bar\nu}_\ell)$ and $W^- \to \ell^{\prime -}
{\bar\nu}_{\ell'}$, with $\ell = e$ and $\ell' = \mu$ or vice-versa.
The background process is therefore $p p \to t {\bar b} \, e^- \mu^-
+~{\hbox{missing transverse energy}}$, to be compared with our process
$p p \to t {\bar b} \, e^- \mu^-$. Using {\tt MadGraph}, we find that
the cross section for this background process and its anti-process at
the 100 TeV FCC-hh is $\sim 60$ fb. With a luminosity of $30~{\rm
  ab}^{-1}$, this corresponds to $1.8 \times 10^6$ events, to be
compared with our $1.7 \times 10^4$ signal events. Using $S/\sqrt{B}$
as a measure of significance ($S$ and $B$ are the number of signal and
background events, respectively), we find $S/\sqrt{B} = 12.7$, which
is excellent. And if cuts on the missing transverse energy are
applied, the background can be considerably reduced.

Second, in order to construct the LITP, it is necessary the measure
the 4-momenta of all the final-state particles. For the $e^-$ and
$\mu^-$, this is no problem. However, how well can $p_t$ and $p_{{\bar
    b}}$ be measured?

We begin with $p_t$. The dominant decay is $t \to b W^+$, with the
$W^+$ decaying to $q {\bar q}$ or $\ell^+ \nu_l$. In either case,
$p_W$ can be found (even though the neutrino is not detected in $W^+
\to \ell^+ \nu_l$, its 4-momentum can be deduced from the constraint
that $p_W^2 = M_W^2$). Similarly, $p_B$ can be deduced from $(p_B +
p_W)^2 = m_t^2$, so that $p_t$ can be obtained. The only potential
difficulty is that there is an ambiguity due to the fact that one
cannot distinguish the $b$ from $t$ decay from the ${\bar b}$ in the
final state. However, similar problems arise when the $t$-quark mass
is measured in a $t{\bar t}$ final state. In this case, there are
statistical methods to deal with the ambiguity, so that the
measurement can be made, with some smearing of the result
\cite{JFthanks}.

The measurement of $p_b$ is more challenging. Roughly 30\% of $b$
decays are semileptonic and include (undetected) neutrinos, which
impacts the precision with which $p_b$ can be measured. Also, for jets
in general, the precision of the measurement of the 4-momenta improves
with larger $p_T$, suggesting that it may be more difficult to measure
$p_b$ as the LHC energy increases. On the other hand, $b$-jet
resolution is extremely important for the ATLAS and CMS Collaborations
in order to be able to observe $H \to b{\bar b}$, so that there is a
good deal of work in this area. For example, it has been found that it
is possible to correct the $b$-jet energy when a muon is found inside
the jet \cite{JFthanks}.

It is clear that a full Monte Carlo simulation is required to
determine how well the LITP signal of Table I can be measured.
However, there are two additional points. First, the $H^-$-${\tilde
  R}_2$-$S_1$ coupling of the toy model has dimensions of a mass, with
$M' = O({\rm TeV})$. The LITP events were generated taking $M' = 1$
TeV. But another value of $M'$ could have been reasonably chosen, say
$M' = 2$ TeV. Since the cross section is proportional to ${M'}^2$,
this would increase the number of events in Table I by 4. Thus, rather
than asking how well the LITP signal of Table I can be measured, a
better question would be: what size of LITP can be determined to, say,
$3\sigma$? Second, while we have focused on ${\bar u} d \to t {\bar b}
\, e^- \mu^-$, other processes are also possible: ${\bar u} d \to t \,
jet \, \, e^- \mu^-$, ${\bar u} d \to jet \, {\bar b} \, e^- \mu^-$
and ${\bar u} d \to jet \, jet \, \, e^- \mu^-$, where $jet$
represents a light quark. LITPs should be searched for in all of these
processes.

\section{Conclusions}

If the neutrino is a Majorana particle, this means that
lepton-number-violating (LNV) processes are possible. These typically
contain a pair of same-sign leptons in the final state.  At low
energies, there are experiments looking for neutrinoless double-beta
($0\nu\beta\beta$) decay, $n n \to p p \, e^- e^-$, or $d d \to u u \,
e^- e^-$ at the quark level. It is also possible to search for LNV
processes at high energies, at the LHC. One advantage at the LHC is
that there are many LNV processes, including those in which the
final-state leptons have different flavors, i.e., there is also lepton
flavor violation. We refer to all of these as $0\nu\beta\beta$-like
processes.

There is also an important disadvantage: the $0\nu\beta\beta$ decay
amplitude is suppressed by a light neutrino mass. If such a
suppression were present in LHC processes, they would be
unobservable. Fortunately, there are many NP models in which
$0\nu\beta\beta$-like processes can be generated without the amplitude
being suppressed by a light neutrino mass. If such a process were
observed at the LHC, it would imply that one of these NP models is
present. Can we figure out which one?

In this paper, we use an effective-field-theory analysis to show that,
if certain pairs of NP operators contribute to a $0\nu\beta\beta$-like
process, when one squares the total amplitude, their interference
generates a term of the form $\epsilon_{\mu \nu \rho \sigma} \,
p_1^\mu p_2^\nu p_3^\rho p_4^\sigma$, where the $p_i$ are the
4-momenta of the final-state particles. This is a CP-violating
Lorentz-invariant, triple product (LITP). In order to illustrate this,
we focus on ${\bar u} d \to t {\bar b} \, e^- \mu^-$.  We construct a
toy model involving a charged Higgs ($H^-$) and two types of
leptoquark (${\tilde R}_2$ and $S_1$), all with masses of 1 TeV.
There is a $H^-$-${\tilde R}_2$-$S_1$ coupling that violates $L$ by
two units and leads to the LNV process ${\bar u} d \to t {\bar b} \,
e^- \mu^-$. The couplings are chosen such that a LITP is generated.

Using {\tt FeynRules} \cite{Alloul:2013bka} and {\tt MadGraph}
\cite{Alwall:2014hca}, we examined the prospects for measuring the
LITP at the LHC. To be specific, we considered the HL-LHC (14 TeV),
the HE-LHC (27 TeV) and the FCC-hh (100 TeV). For each machine, we
generated the ${\bar u} d \to t {\bar b} \, e^- \mu^-$ events. For
each set, we calculated the TP asymmetry $A_{TP}$, defined as the
difference of the percentage of events with LITP $>0$ and LITP $<0$.
If $A_{TP} \ne 0$, this is a signal of CP violation. We find that the
predicted $A_{TP}$ is not measurable at the HL-LHC, may be measurable
at the HE-LHC, and is certainly measurable at the FCC-hh.

What would we learn from such a measurement? This depends on what we
already know at the time of the measurement. If no NP particles have
been found via direct production, the observation of $p p \to t {\bar
  b} \, e^- \mu^-$ would be an indirect confirmation of the presence
of NP, and the measurement of a nonzero TP asymmetry would indicate
that there are two interfering amplitudes with different Lorentz
structures and a nonzero weak-phase difference. And if NP particles
have already been found, this will provide information about their
properties. In the case of the above model, with masses of 1 TeV, it
is likely that the $H^-$, ${\tilde R}_2$ and $S_1$ will already have
been discovered, and some of their couplings to ordinary particles
measured. But it may not be known that there is a $H^-$-${\tilde
  R}_2$-$S_1$ coupling that has $\Delta L=2$. And the measurement of
the TP asymmetry gives phase information about the NP couplings that
would be difficult to obtain otherwise.

\bigskip
{\bf Acknowledgments}: We thank B. Kayser for collaboration in the
early stages of this project, and J.-F. Arguin, B. Bhattacharya,
A. Datta \& J. De Vries for useful discussions. This work was
financially supported by NSERC of Canada.

\end{document}